\documentclass[aps,preprint,onecolumn,showpacs]{revtex4}%
\usepackage{amsfonts}
\usepackage{amsmath}
\usepackage{amssymb}
\usepackage{graphicx}%
\begin{document}
\title{Generation of $N$-qubit $W$ state with rf-SQUID qubits by adiabatic passage}
\author{Z. J. Deng$^{1,2}$ }
\author{K. L. Gao$^{1}$}
\email{klgao@wipm.ac.cn}
\author{M. Feng$^{1}$}
\email{mangfeng1968@yahoo.com}
\affiliation{$^{1}$State Key Laboratory of Magnetic Resonance and Atomic and Molecular
Physics, Wuhan Institute of Physics and Mathematics, Chinese Academy of
Sciences, Wuhan 430071, China}
\affiliation{Centre for Cold Atom Physics, Chinese Academy of Sciences, Wuhan 430071, China}
\affiliation{$^{2}$Graduate School of the Chinese Academy of Sciences, Beijing 100049, China}

\begin{abstract}
A simple scheme is presented to generate $n$-qubit $W$ state with
rf-superconducting quantum interference devices (rf-SQUIDs) in cavity QED
through adiabatic passage. Because of the achievable strong coupling for
rf-SQUID qubits embedded in cavity QED, we can get the desired state with high
success probability. Furthermore, the scheme is insensitive to position
inaccuracy of the rf-SQUIDs. The numerical simulation shows that, by using
present experimental techniques, we can achieve our scheme with very high
success probability, and the fidelity could be eventually unity with the help
of dissipation.

\end{abstract}

\maketitle

Besides being an essential resource for testing quantum mechanics against
local hidden theory \cite{1}, entanglement also has many applications in
quantum information processing (QIP) such as quantum computation \cite{2},
quantum cryptography \cite{3}, quantum teleportation \cite{4} and so on.
Recently, five-photon entanglement \cite{5}, six-ion
Greenberger-Horne-Zeilinger (GHZ) state \cite{6} and eight-ion $W$ state
\cite{7} have been experimentally reported. The $W$ state, whose general form
is $W_{N}=\frac{1}{\sqrt{N}}|N-1,1\rangle$ with $|N-1,1\rangle$ being all the
totally symmetric states involving $N-1$ zeros and $1$ one, is famous for its
entanglement robustness to local operation, even under qubit loss. As far as
we know, while there are some theoretical proposals discussing the generation
of $W$ states in cavity QED using atoms \cite{8,9,10,11}, quantum dots
\cite{12}, and rf-SQUIDs \cite{13,14}, we have not yet found any experimental
report of the $W$ state in cavity QED. In Refs. \cite{8,9}, atoms are required
to pass through the cavity with a certain velocity or velocity distribution,
which is experimentally challenging. Measurement of leaky photons \cite{10,15}
can also lead to multi-atom entanglement, but it is only probabilistic to get
the desired state in one trial and the measurement requires highly efficient
detector. Ref. \cite{13} proposed to generate multi-rf-SQUID $W$ states via
nonresonant interaction with the cavity along with auxiliary measurement,
which evidently restricts the speed and success probability. In Ref.
\cite{14}, by Raman transition, the $W$ state can be efficiently generated.
However, like most of the schemes mentioned above, it requires precise control
of the interaction time.

On the other hand, based on Josephson junction, superconducting qubits
including charge qubits \cite{16,17}, flux qubits \cite{17,18}, and phase
qubits \cite{19}, have attracted much attention due to their potential for
scalability. The rf-SQUID, which consists of an enclosed superconducting loop
interrupted by a Josephson junction, is the simplest design for flux qubits.
When rf-SQUIDs are embedded in a cavity, the strong coupling limit of the
cavity QED, i.e., $g^{2}/(\Gamma\kappa)\gg1$ (parameters explained later),
which is difficult to achieve with atoms in cavity, can be easily realized
\cite{20}. In this paper, we show how to generate multi-qubit $W$ states with
rf-SQUIDs in a cavity by adiabatic passage \cite{20,21}. The biggest merit of
using adiabatic passage is that it does not need precise control of the Rabi
frequency and pulse duration, and the population in excited states is
negligible due to their absence in the dark state used for evolution. Our
scheme has the following advantages: (i) The rf-SQUIDs are fixed in the
cavity, which lowers the experimental difficulty for control comparing to
previous schemes \cite{8,9,22} with atoms flying through cavities or to
schemes \cite{23} that require the trapped atoms to be well localized; (ii) By
virtue of adiabatic passage, there is no need for precise control of the Rabi
frequency and the pulse duration; (iii) By adding the microwave pulse
collinearlly with the cavity mode, the scheme is insensitive to position
inaccuracy of the rf-SQUIDs; (iv) With present experimental technology, the
success probability is very high and the fidelity is eventually unity with the
help of dissipation; (v) Qubits are encoded in the two lowest flux states. So
once it is prepared, the $W$ state can remain in the cavity almost without
energy relaxation.

Let us consider N rf-SQUIDs, each of which has a $\wedge$-type configuration
formed by two lowest levels and an excited level, as shown in Fig. 1. For a
rf-SQUID $j$, the classical field drives the transition resonantly between the
level $|1\rangle_{j}$ and the level $|e\rangle_{j}$ with Rabi frequency
$\Omega_{j},$ while the cavity field couples resonantly to the level
$|0\rangle_{j}$ and the level $|e\rangle_{j}$\ with coupling constant $g_{j}$.
The Hamiltonian of a rf-SQUID $j$ with junction capacitance $C_{j}$, loop
inductance $L_{j}$ and externally applied biased flux $\Phi_{x,j}$ can be
written in the usual form \cite{17,20,24,25},%

\begin{equation}
H_{s,j}=\frac{Q_{j}^{2}}{2C_{j}}+\frac{(\Phi_{j}-\Phi_{x,j})^{2}}{2L_{j}%
}-E_{J,j}\cos(2\pi\frac{\Phi_{j}}{\Phi_{0}}), \label{1}%
\end{equation}
\ \ \ \ where $\Phi_{j}$ is the magnetic flux threading the loop, $Q_{j}$ is
the charge on the leads, and $E_{J,j}=I_{c,j}\Phi_{0}/2\pi$ is the Josephson
energy with $I_{c,j}$ the critical current of the junction and $\Phi_{0}=h/2e$
the flux quantum. So the expressions of $g_{j}$ and $\Omega_{j}$ are given,
respectively, by \cite{20,25},%

\begin{align}
g_{j}  &  =\frac{1}{L_{j}}\sqrt{\frac{\omega_{c}}{2\mu_{0}\hbar}}\text{ }%
_{j}\langle0|\Phi_{j}|e\rangle_{j}\int_{S_{j}}\mathbf{B}_{c}^{j}%
(\mathbf{r})\cdot d\mathbf{S,}\label{2}\\
\Omega_{j}  &  =\frac{1}{L_{j}\hbar}\text{ }_{j}\langle1|\Phi_{j}|e\rangle
_{j}\int_{S_{j}}\mathbf{B}_{\mu w}^{j}(\mathbf{r,t})\cdot d\mathbf{S,}%
\nonumber
\end{align}
here $\omega_{c}$ is the cavity frequency, $\mathbf{B}_{c}^{j}(\mathbf{r})$
and $\mathbf{B}_{\mu w}^{j}(\mathbf{r})$ are the magnetic component of the
cavity mode and the classical microwave in the superconducting loop of the
$j$th rf-SQUID, respectively. We assume that all the N rf-SQUIDs are
identical. The Hamiltonian of N rf-SQUIDs interacting simultaneously with a
single cavity mode and a microwave pulse in the interaction picture is
described by (assuming $\hbar=1$)%

\begin{equation}
H_{I}=\overset{N}{\underset{j=1}{\sum}}g_{j}(a^{\dagger}|0\rangle_{jj}\langle
e|+a|e\rangle_{jj}\langle0|)+\Omega_{j}(t)(|1\rangle_{jj}\langle
e|+|e\rangle_{jj}\langle1|), \label{3}%
\end{equation}
where $a^{\dagger}$ and $a$ are, respectively, the creation and annihilation
operators for the cavity mode. Initially all the qubits are in the ground
state, i.e., $|0\rangle_{1,2,...,N}$ and the cavity is in the single-photon
state $|1\rangle$. The dark state associated with such an initial state is%

\begin{align}
|D(t)\rangle &  \propto|0\rangle_{1,2,...,N}|1\rangle-(\frac{g_{1}}{\Omega
_{1}(t)}|1\rangle_{1}|0\rangle_{2}...|0\rangle_{N-1}|0\rangle_{N}+\nonumber\\
&  \frac{g_{2}}{\Omega_{2}(t)}|0\rangle_{1}|1\rangle_{2}...|0\rangle
_{N-1}|0\rangle_{N}+...+\frac{g_{N}}{\Omega_{N}(t)}|0\rangle_{1}|0\rangle
_{2}...|0\rangle_{N-1}|1\rangle_{N})|0\rangle.
\end{align}

If the microwave pulse is input collinearlly with the cavity axis and thereby
the driving pulse will have the same spatial mode structure as the cavity mode
\cite{26}, i.e., $\mathbf{B}_{\mu w}^{j}(\mathbf{r,t})\propto\overset{\sim
}{\Omega}(t)\mathbf{B}_{c}^{j}(\mathbf{r})$, with $\overset{\sim}{\Omega}(t)$
being a variable that reflects the time depence of the microwave pulse's
amplitude, then we have $\frac{g_{j}}{\Omega_{j}(t)}=\frac{K}{\overset{\sim
}{\Omega}(t)}$ ($j=1,2,...,N$ and $K$ is a constant) and Eq. (4) reduces to%

\begin{equation}
|D(t)\rangle\propto|0\rangle_{1,2,...,N}|1\rangle-\frac{K\sqrt{N}}%
{\overset{\sim}{\Omega}(t)}W_{N}|0\rangle. \label{5}%
\end{equation}

By reducing $\overset{\sim}{\Omega}(t)$ adiabatically, we can get the
$N$-qubit $W$ state $W_{N}$ without the requirement of accurate positions of
the rf-SQUIDs. To carry out our scheme, we first need to prepare the
single-photon cavity state, which can be done as follows: Initially the cavity
is in the vacuum state $|0\rangle$ and the first rf-SQUID is in state
$|1\rangle_{1}$, while the other $N-1$ rf-SQUID qubits are in state
$|0\rangle_{2,3...,N}$. After simultaneously adjusting, by the externally
applied biased flux $\Phi_{x,j},$ the level spacing of the rf-SQUID qubits,
from the second to the Nth, to be decoupled from the cavity mode and the
microwave, we drive $|1\rangle_{1}|0\rangle,$ by a microwave with increasing
amplitude, to evolve along the dark state $g_{1}|1\rangle_{1}|0\rangle-$
$\Omega_{1}(t)|0\rangle_{1}|1\rangle$ to the single-photon cavity state and
the state $|0\rangle_{1}$\ of the first rf-SQUID. Then by adjusting the biased
flux again, we bring back the level spacing of the other $N-1$ rf-SQUID qubits
simultaneously to the original situation.

To check the feasibility of our scheme, we consider below the decaying effects
on our W state preparation. To this end, we will numerically simulate the
system evolution by quantum jump approach \cite{27} under the condition that
no photon leakage actually happens, due to either the spontaneous emission
from the rf-SQUID's excited state or the decay from the cavity mode, during
the preparation period. In real experiments, the values of $g_{j}$ and
$\Omega_{j}(t)$ may vary from qubit to qubit, but their ratio is a constant as
analysed above, which guarantees that it has no effect on the fidelity of the
generated $W$ state. For simplicity, we assume all the qubits have the same
coupling strengths, i.e., $g_{j}=g$, $\Omega_{j}(t)=\Omega(t)$ ($j=1,2,...,N$%
). Similar to Ref. \cite{20}, we suppose $\Omega(t)=g\times40\times\exp
[-t^{2}/2\tau^{2}]$ with $\tau=4/g$ and choose the coupling $g=1.8\times
10^{8}$ s$^{-1}$, the spontaneous emission rate of the excited level
$\Gamma=4\times10^{5}$ s$^{-1}$, and the cavity decay rate $\kappa
=1.32\times10^{6}$ s$^{-1},$ which are available with the present experimental
technology \cite{20}. In Fig. 2, we show the time evolution of the three-qubit
case. In Fig. 2 (a), we find small oscillations of the unwanted states. By
enlarging $\Gamma$ to $4\times10^{7}$ s$^{-1}$ in Fig. 2 (b), we can much
suppress those small oscillations. This is because larger spontaneous emission
rate can damp the population of unwanted states faster, so the fidelity in (b)
at the time $t=25/g$ is F=0.9994, higher than that in (a) with F=0.9946, at
the cost of a lower success probability at that time point. Obviously, with
the help of dissipation, the fidelity is eventually unity. We have also
plotted Fig. 3 to show how our scheme works with the multi-qubit case: For a
certain qubit number, the longer the evolution time, the higher the fidelity
and the lower the success probability, and eventually, the fidelity could be
unity and the success probability remains a constant. We show in Fig. 3 that
the fidelities are almost unity at time $t=50/g$ for the qubit number ranging
from 3 to 80, and the corresponding success probabilities are all higher than
90.5\%. The optimal success probability and fidelity for a certain qubit
number can be obtained by changing the pulse maximum height and pulse duration
time. For simplicity, we have assumed in our numerical simulation the same
pulse for different qubit number cases, because we just want to illustrate
that our scheme could work well in the case of different qubit number.

As mentioned above, there have been some proposals for preparing $W$ state in
cavity QED. For example, in Refs. \cite{8,9}, atoms need to be well controlled
with velocity in passing through the cavity due to the requirement of the
interaction time, which is experimentally very difficult. In contrast, SQUIDs
are static and controllable in our scheme. Ref. \cite{11} also generated the
multi-atom $W$ state by adiabatic passage. But due to the small value of
$g^{2}/(\Gamma\kappa)$ regarding the atom-cavity system, the success
probability is low. Moreover, as the scheme starts from the initial state,
i.e., $|1\rangle_{1}|0\rangle_{2,3...,N}|0\rangle$, the timescale for the
population transfer is ten times larger than that in our scheme \cite{28},
which further reduces the success probability. Refs. \cite{10,13} require
auxiliary measurement, so the success probability would also be low. In
contrast, our scheme can be reached with high success probability because of
the achievable strong coupling limit of cavity QED with the rf-SQUIDs. Thus we
can reduce the repetitional times in experiments. Furthermore, by Raman
transition, the W state can also be generated with high success probability in
Ref. \cite{14}, based on the requirements of precise control of the pulse
duration time and of accurate positions of the SQUIDs. However, in real
experiments, these requirements are hard to be met. In this sense, our scheme
is much advantageous with the adiabatic passage and the driving pulse added
collinearly with the cavity mode.

In summary, we have presented a simple scheme of generating $N$-qubit $W$
state with rf-SQUIDs by adiabatic passage. Our numerical simulation has shown
that our scheme can be achieved with high success probability and unity
fidelity. Moreover, it has some favorable features for experimental
implementation. So we argue that our proposal might be experimentally realized
with present technology and would be useful for experimental implementation of
solid-state quantum information processing.

{\large ACKNOWLEDGMENTS}

Z. J. Deng is grateful to Jie-Qiao Liao, Xiong-Jun Liu and X. L. Zhang for
their warmhearted helps. This work is partly supported by National Natural
Science Foundation of China under Grant Nos. 10474118 and 60490280, by Hubei
Provincial Funding for Distinguished Young Scholars, and partly by the
National Fundamental Research Program of China under Grant No. 2005CB724502.


\begin{thebibliography}{99}
\bibitem {28}It is found by our simulation from the initial state
$|1\rangle_{1}|0\rangle_{2,3...,N}|0\rangle$ with the same parameters as in
Fig. 2 (b).\newpage

{\LARGE Figure captions}

FIG. 1. The level configuration for the $j$th rf-SQUID, where level
$|1\rangle_{j}$ and the level $|e\rangle_{j}$ are coupled resonantly by the
classical microwave, while level $|0\rangle_{j}$ and the level $|e\rangle_{j}$
are coupled resonantly by the cavity mode. $\Omega_{j}$ and $g_{j}$ are
coupling strengths regarding the microwave and the cavity mode, respectively.
The qubits are encoded in levels $|0\rangle_{j}$ and $|1\rangle_{j}$.

FIG. 2. Time evolution with respect to the three-qubit case, where
$g=1.8\times10^{8}$ s$^{-1}$, $\kappa=1.32\times10^{6}$ s$^{-1}$ and
$\Gamma=4\times10^{5}$ s$^{-1}$ in (a) and $\Gamma=4\times10^{7}$ s$^{-1}$ in
(b). The dotted, dashed and solid lines represent the populations of
$|0\rangle_{1}|0\rangle_{2}|0\rangle_{3}|1\rangle$, of the $W$ state (i.e.,
$|1\rangle_{1}|0\rangle_{2}|0\rangle_{3}|0\rangle$, $|0\rangle_{1}%
|1\rangle_{2}|0\rangle_{3}|0\rangle$, $|0\rangle_{1}|0\rangle_{2}|1\rangle
_{3}|0\rangle$) and of the states containing the excited state $|e\rangle_{j}$
(i.e., $|e\rangle_{1}|0\rangle_{2}|0\rangle_{3}|0\rangle$, $|0\rangle
_{1}|e\rangle_{2}|0\rangle_{3}|0\rangle$, $|0\rangle_{1}|0\rangle_{2}%
|e\rangle_{3}|0\rangle$), respectively, while the dashed-dotted line denotes
the success probability P(t). The values of F and P labeled in the plot are
the fidelity and success probability at time $t=25/g$.

FIG. 3. Success probability P and fidelity F versus the qubit number N, where
$g=1.8\times10^{8}$ s$^{-1}$, $\kappa=1.32\times10^{6}$ s$^{-1}$ and
$\Gamma=4\times10^{7}$ s$^{-1}$. The circle and five-pointed star denote the
fidelities at time $t=25/g$ and $50/g,$ respectively, and the plus sign and
the asterisk are their corresponding success probabilities.
\end{thebibliography}
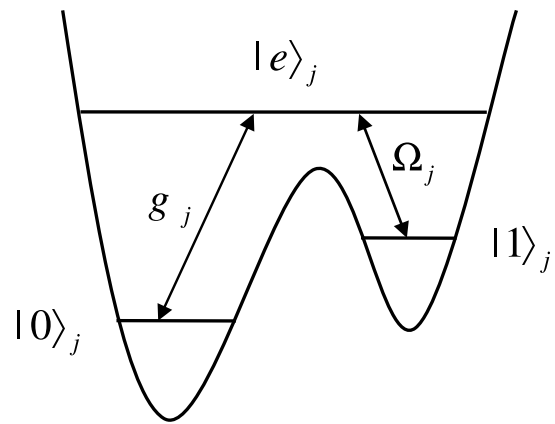
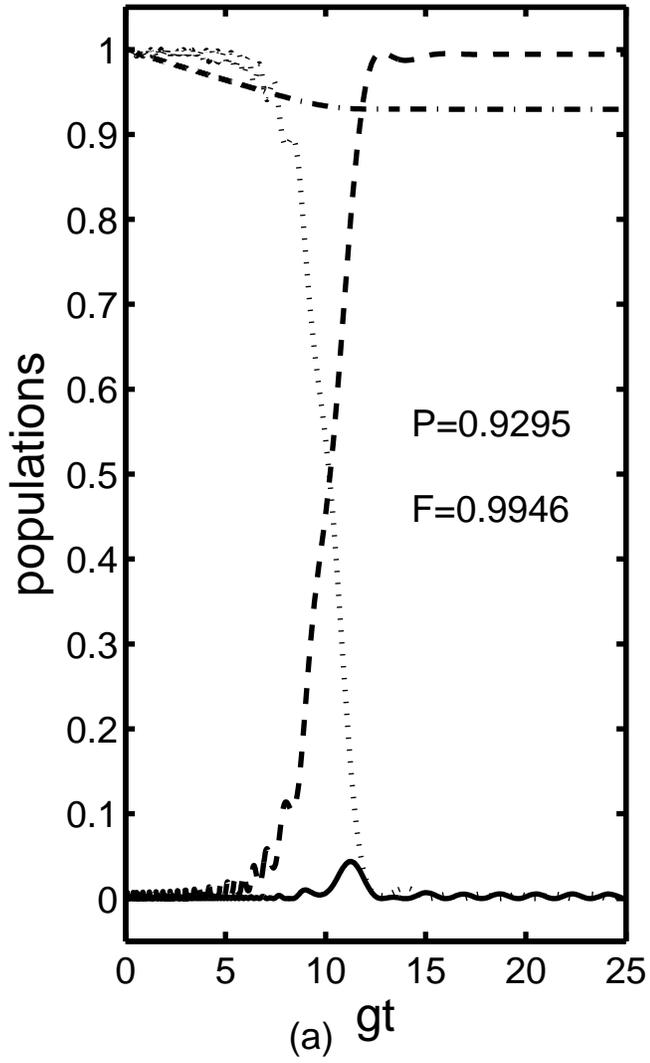
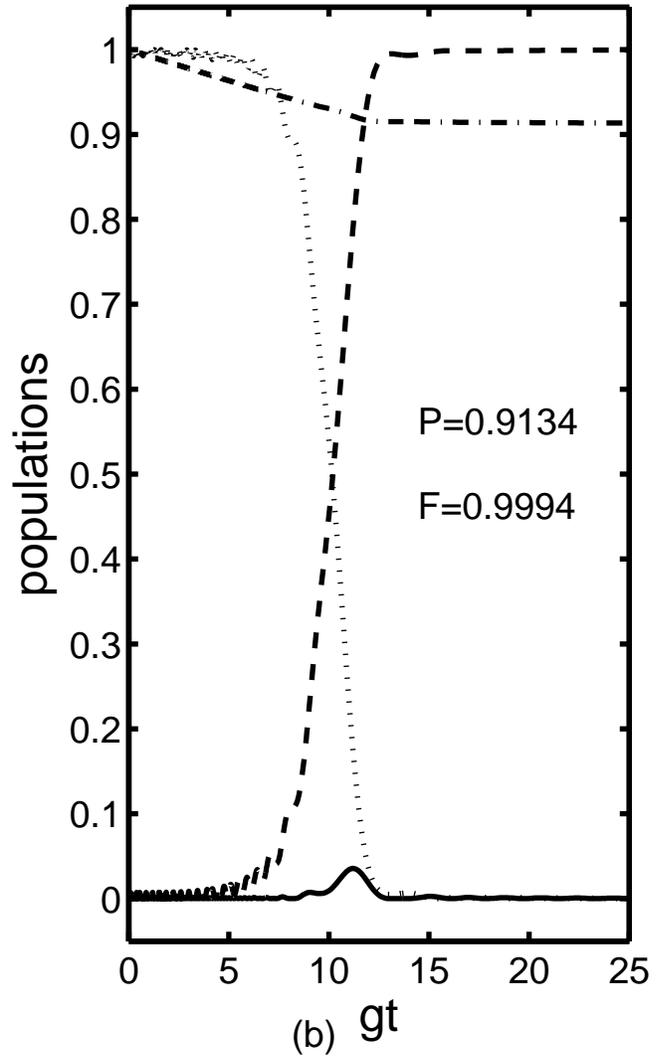
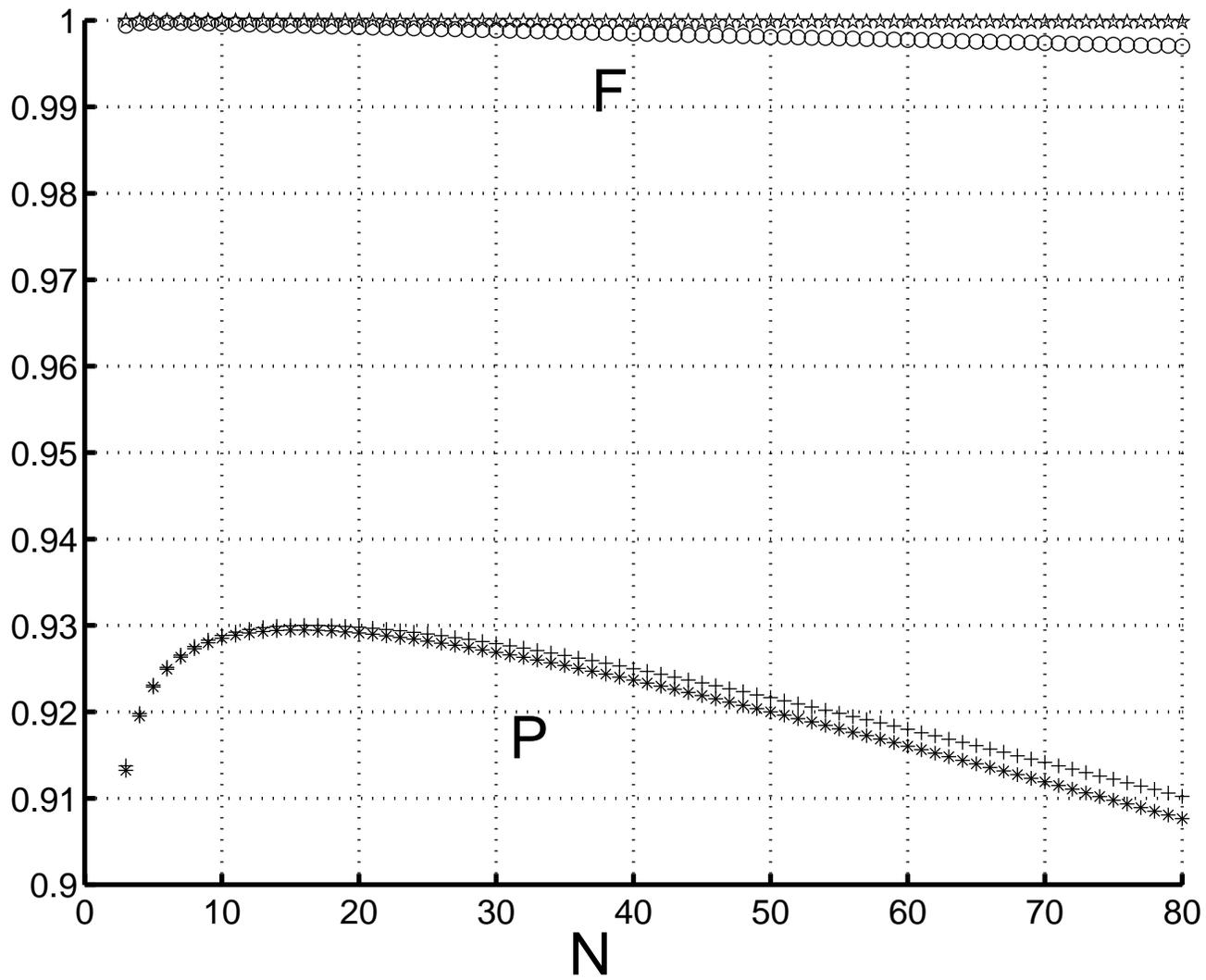
\end{document}